# "The Human Body is a Black Box": Supporting Clinical Decision-Making with Deep Learning

Mark Sendak
Duke Institute for Health Innovation
Durham NC USA
mark.sendak@duke.edu

Madeleine Clare Elish
Data & Society Research Institute
New York NY USA
mcelish@datasociety.net

Michael Gao
Duke Institute for Health Innovation
Durham NC USA
michael.gao@duke.edu

Joseph Futoma[†]
Engineering & Applied Sciences
Harvard University
Cambridge MA USA
jfutoma@seas.harvard.edu

William Ratliff
Duke Institute for Health Innovation
Durham NC USA
william.ratliff@duke.edu

Marshall Nichols
Duke Institute for Health Innovation
Durham NC USA
marshall.nichols@duke.edu

Armando Bedoya
Pulmonology and Critical Care
Duke School of Medicine
Durham NC USA
armando.bedoya@duke.edu

Suresh Balu
Duke Institute for Health Innovation
Durham NC USA
suresh.balu@duke.edu

Cara O'Brien
Hospital Medicine
Duke School of Medicine
Durham NC USA
cara.obrien@duke.edu

## ABSTRACT

Machine learning technologies are increasingly developed for use in healthcare. While research communities have focused on creating state-of-the-art models, there has been less focus on real world implementation and the associated challenges to fairness, transparency, and accountability that come from actual, situated use. Serious questions remain underexamined regarding how to ethically build models, interpret and explain model output, recognize and account for biases, and minimize disruptions to professional expertise and work cultures. We address this gap in the literature and provide a detailed case study covering the development, implementation, and evaluation of Sepsis Watch, a machine learning-driven tool that assists hospital clinicians in the early diagnosis and treatment of sepsis. Sepsis is a severe infection that can lead to organ failure or death if not treated in time and is the leading cause of inpatient deaths in US hospitals. We, the team that developed and evaluated the tool, discuss our conceptualization of the tool not as a model deployed in the world but instead as a socio-technical system requiring integration into existing social and professional contexts. Rather than focusing solely on model interpretability to ensure fair and accountable machine learning, we point toward four key values and practices that should be considered when developing machine learning to support clinical decision-making: rigorously define the problem in context, build relationships with stakeholders, respect professional discretion, and create ongoing feedback loops with stakeholders. Our work has significant implications for future research regarding mechanisms of institutional accountability and considerations for responsibly designing machine learning systems. Our work underscores the limits of model interpretability as a solution to ensure transparency, accuracy, and accountability in practice. Instead, our work demonstrates other means and goals to achieve FATML values in design and in practice.

## 1 INTRODUCTION

Machine learning technologies are increasingly developed for use in healthcare. From consumer facing apps to hospital readmission predictors, the healthcare industry includes a rapidly expanding set of use cases for machine learning applications [59]. The machine learning community has focused much research on creating state-of-the-art models, but there has been less focus on real world implementation and the associated challenges to fairness, transparency, and accountability that come from actual, situated use. Serious questions remain underexamined regarding how to ethically build models, interpret and explain model output, recognize and account for biases, and minimize disruptions to professional expertise and work cultures.

This paper contributes a case study through which to examine how issues of transparency, trust, and accountability are grappled with in practice. We present an empirical case study covering the development, implementation, and evaluation of Sepsis Watch, a machine learning-driven tool that assists hospital clinicians in the early diagnosis and treatment of sepsis. Sepsis is an inflammatory response to infection that can lead to organ failure and is the leading cause of inpatient deaths in US hospital [47]. We, the authors, were part of the team developing and evaluating this tool and our case study is based on our practitioner and research experiences.

Some of the unique challenges facing the development of Sepsis Watch were that sepsis is not only hard to predict but also lacks a universally accepted definition. A model to predict sepsis needed to articulate a ground truth where in fact there was none. Moreover, when and why sepsis develops is incompletely understood. These aspects contributed to a de-prioritization of model interpretability in favor of other ways to establish trust in the accuracy of the model with clinicians, including rigorous documentation and institution-specific validation and evaluation.

At the same time, our interdisciplinary team approached the development of Sepsis Watch as the development of a socio-technical system, not an isolated model. By using the term "socio-technical system," we mean to foreground the interconnected social and technical dimensions of a technology, in which, for instance, the use of a machine learning tool cannot be considered apart from the people and institutions who interact with the tool and the beliefs, contexts, and power hierarchies that shape its development and use. From this perspective, Sepsis Watch is a great deal more than a deep learning model that generates risk scores; it is a complex socio-technical system that combines technical and institutional infrastructures with professionals who are making critical and highly contextual decisions.

In this paper, we present our approach and describe the processes of designing, developing, and implementing Sepsis Watch. We begin by situating our analysis in related work on machine learning healthcare applications and the literature on technology adoption in healthcare institutions. This is followed by an overview of the Sepsis Watch tool and a discussion of our work building trust and mechanisms of accountability at multiple levels of the project, over time and with different stakeholders. We employ the term trust to describe a belief held by individuals that the system in place is appropriate and accurate, and accountability to refer to the ways in which technologists and designers can be held responsible for the performance of the system. We also discuss the complexities of integrating the output of the model into clinical decision-making. Drawing on these experiences and observations, we conclude by presenting four key values and practices that should be considered when developing machine learning to support clinical decision-making: rigorously define the problem in context, build relationships with stakeholders, respect professional discretion, and create ongoing feedback loops with stakeholders.

Our contribution to the growing literature on fair, accountable, and transparent machine learning (FATML) is two-fold. First, we provide a detailed case study of a tool's development and implementation, presenting empirical and socially-situated (as opposed to experimental) evidence of a tool in use and its intersections with existing FATML concerns. To our knowledge, Sepsis Watch is comparable to no other clinical decision support system, and is one of the first deep learning models to be fully integrated into routine clinical care. Second, we demonstrate the implications for trust, accountability, and transparency when a machine learning implementation is conceptualized not as a model deployed in the world but rather as a socio-technical system that must be integrated into existing social and professional contexts. From this perspective, model interpretability itself does not ensure a fair or accountable technology. Rather, our experience and discussion points toward significant alternative mechanisms to achieve FATML values in design and in practice.

## 2 MACHINE LEARNING FOR HEALTHCARE

The introduction of machine learning systems is often imagined as a profound disruption, leading to either a wholly transformed system or the demise of human expertise. In healthcare, technology implementation and adoption decisions are made at the leadership level and depend on regulatory and compliance oversight [23, 44]. There is limited literature on the nuances of how machine learning models challenge the ways healthcare organizations function and clinical professionals relate to their work.

Debates about the introduction of new technologies into existing clinical care routines are far from new. There is much to be gained from understanding how previous technologies were introduced into healthcare settings, how they were perceived when they were new, and why some innovations are more successful than others. At the same time, big data and machine learning rightly seem to present unique challenges with consequences for fairness, transparency, and accountability that require new analyses and governance principles [6, 10, 29].

### 2.1 IS INTERPRETABILITY NECESSARY?

Both researchers and practitioners have highlighted the potential opportunities and perils of machine learning for healthcare. Numerous reviews document the proliferation of machine learning models and the promise of these technologies to address challenges in healthcare [45, 59]. At the same time, concerns have been raised about the role machine learning technologies might play in exacerbating systemic inequities in healthcare, undercutting patient privacy, increasing surveillance of vulnerable populations, and "dehumanizing" medicine [17, 38, 41, 60].

The need for interpretable or explainable machine learning has taken hold in the context of these concerns, as well as in the context of facilitating and regulating adoption. Many experts cite explainable machine learning as the answer to protecting patient safety and ensuring professional accountability [24, 49]. The US National Artificial Intelligence Research and Development Strategic Plan recently stated, "Truly trustworthy AI requires explainable AI... this requires a comprehensive understanding of the AI system by the human user and the human designer" [52]. While "trust" is not synonymous with adoption, in a highly regulated, risk-averse industry such as healthcare, "trust" is often necessary in order for adoption to occur.

Several assumptions underlie the focus on explainability as the primary means by which to address concerns around accountability, transparency, trust, and adoption of machine learning in healthcare. First, professional clinicians are presumed to have substantial technical and quantitative expertise with which to engage with explainable machine learning and "to interrogate, understand, debug and even improve the machine learning system" [2]. A potential by-product of this engagement often left undiscussed is whether physicians want to be increasingly oriented towards technologies rather than patients. Many clinicians in fact hope for machine learning to level the field of medical knowledge and "lead to a new premium: to find and train doctors who have the highest level of emotional intelligence" [58].

Second, professional clinicians often incorporate information into clinical decisions without a comprehensive understanding of the mechanism by which the information is generated. For example, specialized clinical pathologists analyze laboratory specimens such as blood to generate results that are utilized by non-specialist clinicians. Although non-specialist clinicians understand laboratory test operating characteristics such as sensitivity and specificity, they often lack insight into how exactly laboratory tests function. Regulatory regimes ensure that laboratory tests are reliable and promote trust amongst non-specialist clinicians. In addition, as Atul Gawande warned newly minted physicians: "the volume and complexity of the knowledge that we need to master has grown exponentially beyond our capacity as individuals" [20]. Physicians who share this view prioritize the effective use of information in clinical decision making rather than comprehensive understanding of how information is generated.

Moreover, the application of medical knowledge does not necessarily require the identification of causal relations. The human body is in many ways "a black box," in which the causes and mechanisms of illnesses often elude explanation. In the case of sepsis, as we discuss in further detail below, this is particularly relevant. What should constitute an explainable algorithm in clinical practice when the definition and underlying pathophysiology of sepsis are incompletely understood in the first place?

Finally, there remains a lack of consensus around what should or could constitute "interpretable" or "explainable" AI [13, 35, 37, 61]. Recent work has emphasized that explanations cannot be an end in and of themselves, but must be enacted in the service of specific normative ends within broader social contexts [26, 51].

## 2.2 ADOPTION OF HEALTH INNOVATIONS

The challenges facing the introduction of machine learning into healthcare settings are an increasing point of discussion both in academic and popular literature. However, most discussions of machine learning adoption fail to place the current challenges in the context of broader innovation adoption theories. The actual mechanisms by which professionals trust, adopt, and use technologies are often an afterthought. Clinicians are rarely directly involved in the development of machine learning technologies [54] or discussions of explainability in healthcare [31]. Too often, the focus has been on the technical properties of the model, and in turn, the potential solutions are intrinsically technical. In contrast, innovation adoption theory focuses on the dimensions of technology use in organizational and professional contexts.

Two features of innovative products that are highlighted across innovation adoption frameworks are perceived benefit and ease of use. The Technology Acceptance Model (TAM) posits that perceived utility and ease of use shape attitudes towards an innovation, which in turn translates into intention to use and actual use [25]. The second framework, modeled after Rogers' "Diffusion of Innovations", extends these two properties to also include alignment with values and experiences of end users, ability to test an innovation, and the ability to watch others try an innovation first [5, 48]. In both frameworks, ease of use is distinct from comprehensive explainability. For example, Intermountain Health Care reduced the rate of pressure sores by 80% by distilling 30-pages of guidelines into 2 simple, high-yield interventions [5]. Explanations may be retrievable, but are not often used in routine clinical practice.

Beyond properties of individual products, two frameworks emphasize the socio-technical dimensions of innovation adoption. John Kotter's "Eight Steps to Leading Change" includes social challenges such as forming a powerful guiding coalition, communicating a vision, empowering broad based action, and changing organizational culture by institutionalizing new approaches [28]. Trish Greenhalgh's "Nonadoption, Abandonment, Scale-up, Spread, and Sustainability" (NASSS) framework consists of 6 domains other than the technology at hand, including the wider political and regulatory system and organizational characteristics such as leadership and readiness for change [22]. Both frameworks emphasize the role of human labor required to integrate rather than deploy machine learning technologies within complex organizations [34].

Although the frameworks described above provide a foundation of important insights into how healthcare innovations are adopted, machine learning technologies present a new set of challenges. New mechanisms of trust and accountability must be developed with both hospital leaders and front-line staff to incorporate this new type of prediction and information into clinical practice. In addition, regulatory frameworks for machine learning in healthcare are nascent, requiring substantial effort at the local level to build trust and accountability.

## 3 CASE STUDY: SEPSIS WATCH

In this case study, we begin by providing an overview of sepsis and the Sepsis Watch system, including phases of design, development, and workflow. In the second half of the case study, we articulate how we conceptualized Sepsis Watch as a socio-technical system within a particular institutional context. We draw out the implications of this conceptualization by describing a set of strategies used to build trust and mechanisms of accountability around the tool. We end the case study with a discussion of the unanticipated ways in which Sepsis Watch was integrated into and justified within clinical decision-making.

Described in further detail below, Sepsis Watch is a sepsis detection and management platform developed to support and improve patient outcomes through increased compliance with recommended treatment guidelines for sepsis. The technical components of Sepsis Watch include a deep learning model, web application, data pipeline and database, and custom web services to extract data in real-time. When a patient arrives and is admitted to the Emergency Department, her personal electronic health record (EHR) data is run through the Sepsis Watch system. If the model predicts that she is at high risk of developing sepsis, her patient information is represented by a "patient card" on the Sepsis Watch iPad application. A nurse, who is responsible for monitoring the devoted iPad on which Sepsis Watch runs, regularly checks the app to review patients at risk of developing sepsis. If a patient is predicted to be septic or at high risk, the nurse calls the Emergency Department physician responsible for the patient's care, and conveys the risk category on the telephone. If the physician agrees the patient requires treatment for sepsis, the patient is further tracked on the iPad application by the nurse until the recommended treatment for sepsis is completed. This overview of the Sepsis Watch system represents only the most basic aspects of the system. Our case study fleshes out all the human components and technical infrastructure that were put in place to effectively integrate Sepsis Watch into clinical practice.

This article represents an interdisciplinary effort to analyze and articulate the implementation of a deep learning tool into routine clinical healthcare, combining practitioner self-reflection with qualitative insights drawn from ethnographic interviews and observations. The case study draws on practitioner, clinician, and researcher experiences over the 2.5-year duration of the Sepsis Watch project. A detailed description of clinical implementation is provided elsewhere [53]. The Sepsis Watch team included nurses, physicians across multiple specialties, informaticians, statisticians, data engineers, solution architects and user interface designers. Advisors with expertise in clinical research and healthcare regulation were regularly consulted.

An independently-funded anthropologist conducted participant observations and interviews to research the impact of Sepsis Watch on professional relationships, the work environment, and how humans interacted with the technology and collaborated to effectively use the technology. Over the course of the project development and implementation, twenty-seven on-site interviews were conducted with representative stakeholders, including clinicians, technologists, and administrators. In addition, on-site clinical practice was investigated during more than thirty hours of observation. Data analysis and coding was conducted using a grounded theory approach [57].

### 3.1 SEPSIS

Sepsis is an abnormal inflammatory response to infection and is the leading cause of inpatient mortality within hospitals in the United States [47]. In our local setting, about one in five patients admitted to the hospital develop sepsis and about one in ten patients who develop sepsis die during the inpatient stay [30].

While sepsis is universally recognized as a major health concern, there is no one standard way to diagnose sepsis. Different sepsis definitions identify different types of patients, and one expert review concluded that "it is an elusive task to generate a single all-encompassing definition" [3]. Most recently, unsupervised machine learning was utilized to identify four sepsis subgroups, resulting in a collection of disease definitions that are not transparent to even expert clinicians [55]. Similarly, when human sepsis experts from across the country were asked to review patient cases, raters often failed to agree on a sepsis diagnosis [16].

Amidst a lack of professional consensus surrounding sepsis diagnosis, there is close regulatory monitoring of sepsis treatment. In 2015, the Centers for Medicare and Medicaid Services (CMS) began requiring US hospitals to report compliance rates with a sepsis treatment protocol, known as SEP-1. CMS assesses the quality of healthcare provided at over 4,000 health systems in the United States and posts the assessments publicly.[1] In addition, every quarter, health systems manually comb through randomly selected charts to document compliance with the SEP-1 sepsis treatment guidelines that have been documented to improve patient outcomes [46, 56]. While some health systems have improved performance and patient outcomes through quality improvement programs [1], other efforts to integrate technology to improve sepsis detection have failed [4, 14]. Treating sepsis according to evidence-based guidelines and improving patient outcomes remains elusive for most hospital systems.

### 3.2 PROJECT BACKGROUND

In 2016, a team of front-line physicians submitted a pilot proposal to work with a local innovation team to improve detection and treatment of sepsis. Proposals are primarily evaluated based on the importance of the problem and opportunity for improvement rather than the novelty of a technology solution; that is, a problem—not a technology—is the starting point of a project. Health system leaders approved the project for investment and implementation support.

This was not the first time the health system attempted to improve its comparatively low sepsis detection and treatment performance. A prior effort caused front-line staff significant alarm fatigue and failed to improve patient outcomes [4].

---

[1] Quality assessments are posted publicly at www.medicare.gov/hospitalcompare.

Incorporating lessons from this prior implementation, a transdiciplinary team was assembled with the primary objective to improve early detection and treatment for sepsis, as defined by the CMS performance measure.

A guiding principle during the period of defining the problem that Sepsis Watch was trying to solve was to "think beyond detection" in the words of one local physician. Clinicians felt that most sepsis treatment failures were due to failure to follow-up rather than failure to detect. Compliance with SEP-1 requires the successful execution of dozens of steps in the six hours following sepsis detection. The bulk of the Sepsis Watch functionality was designed to continuously monitor patient status and support the tracking of sepsis treatment items to ensure appropriate completion. Clinical stakeholders decided to focus the pilot on early sepsis detection within the emergency department (ED). Analysis of local data revealed that over 40% of sepsis cases emerged in the ED and the average time between admission to the hospital and sepsis was 2.01 hours [30]. Thus, the local context and the local expertise of clinicians shaped the problem definition at the outset of the project.

The first twelve months of the project were dedicated to assembling the team around the problem, curating the data to better characterize the problem, and starting to design a workflow and technology solution. The second twelve months of the project were largely dedicated to developing and validating the ML model, web application, data pipeline, and integration platform which are collectively described as Sepsis Watch. The last six months of the project leading up to launch included technical integration into EHR to access real-time data and integration into clinical workflows.

Development of the model and evaluation of Sepsis Watch in clinical practice was approved by the University Health System Institutional Review Board (Pro00093721, Pro00080914). The deep learning model was published and reported according to the Transparent Reporting of a Multivariable Prediction Model for Individual Prognosis or Diagnosis (TRIPOD) guidelines [11]. Clinical impact is being evaluated according to a pre-registered clinical trial (ClinicalTrials.gov ID: NCT03655626).

## 3.3 DESIGN AND DEVELOPMENT

*3.3.1 Datasets.* The first challenge was to curate local EHR data to better understand the problem and to build out the relevant data elements to train the machine learning model. Clinicians felt that a model developed using local data would outperform models developed elsewhere and prior research has demonstrated the poor generalizability of machine learning models in healthcare [42, 62, 63]. Data were curated from the local quarternary academic hospital with over 1,000 beds and over 40,000 inpatient admissions per year. In total, the model development and evaluation dataset contained over 32 million data points.

*3.3.2 Model.* Machine learning experts explored methods that utilized a broad range of features, including medical history and all repeated vital sign and laboratory measurements. Sepsis is a complex condition that progresses rapidly and results in abnormalities across many data modalities. A model would need to generate accurate predictions starting at the beginning of a hospital encounter and rapidly update as new information became available. Model explainability was not prioritized, because regulations promote standardized treatment of sepsis, regardless of cause. As previously described, human experts often disagree about sepsis diagnoses and major medical and public health organizations publicly promote distinct disease definitions.

The model generates risk scores every hour for every adult patient to detect sepsis [18, 19]. The model extended prior work using recurrent neural networks (RNNs) for clinical event detection by coupling a RNN [9, 33] with a multi-task gaussian processes (MGPs). Although there are emerging methods to improve explainability of RNNs, end users cannot contemplate the entire model and cannot reliably understand the relationships between model inputs and outputs [13, 32]. As such, there was not a deliberate effort to explain the MGP-RNN model output.

*3.3.2 Workflow.* Clinical stakeholders proposed a workflow in which a specialized team of nurses, known as rapid response team (RRT) nurses, were the primary end users of Sepsis Watch. A prior implementation of clinical decision support to improve sepsis detection caused significant alarm fatigue for front-line clinicians—a pop-up fired over 100 times per day in the electronic health record (EHR) for certain high risk patients and 86% of notifications were canceled [4]. Based on this experience, the team prioritized minimizing alarm fatigue by having a nurse screen alerts and make a phone call. The RRT nurse would call front-line physicians to confirm sepsis diagnoses and there was a clear directive that the nurse could not independently diagnose or treat sepsis. Sepsis Watch is not a diagnostic device and was never intended to drive clinical care. Sepsis Watch identified patients for further evaluation and the attending physician caring for an individual patient made the final diagnostic determination to start treatment for sepsis.

The Sepsis Watch workflow was designed to rapidly identify patients requiring treatment for sepsis. Risk scores are calculated every hour and patients with a high risk score above 60% are displayed in red. At this risk score threshold, the positive predictive value of the model was 20%. Sepsis criteria are evaluated every five minutes and patients who meet sepsis criteria are displayed in black. Patients meeting sepsis criteria and high risk patients are displayed at the top of the Triage page. The RRT nurse is instructed to call the ED attending physician to discuss every high risk or septic patient. If the attending physician confirms that the patient does not require treatment for sepsis, the patient is moved to the Screened page. If the attending physician confirms that the patient does require treatment for sepsis, the patient is moved to the Treatment page. Patients are only moved to the Screened or Treatment pages after a telephone conversation between the RRT nurse and ED attending physician.

## 3.4 IMPLEMENTATION

*3.4.1 Cultivating Trust and Accountability to Integrate a Socio-Technical System.* Early on, our team recognized the need to integrate Sepsis Watch into existing social and professional contexts within the local healthcare delivery organization. Prior literature would suggest that model explainability and interpretability are central to cultivating trust, the belief held by an individual that the system is appropriate and accurate, and accountability, the ways in which technologists and designers can be held responsible for the performance of a system, in the context of a machine learning technology [24, 49]. In this section, we highlight actual strategies employed by the Sepsis Watch team to successfully cultivate trust and accountability across stakeholder groups that did not involve model interpretability. We first describe how trust and accountability were cultivated in the team of individuals assembled around Sepsis Watch and then in the approaches to design and develop the technology. We then discuss how Sepsis Watch became trustworthy and accountable at scale and how information technology leaders, hospital leaders, and front-line staff were highly engaged in the design, development, and integration of Sepsis Watch into clinical care. In investigating mechanisms of accountability, our focus is not on individual clinicians, but rather on those who designed and managed the implementation of the tool. While there is a strong history and legal precedent for holding individual clinicians accountable for clinical decisions, there is also increasing recognition that health systems must promote safety at a system level to support the practice of individual clinicians [27].

*3.4.2 Team Members Embedded in Existing Social and Professional Networks.* The team assembled to design, develop, and integrate Sepsis Watch into routine clinical practice included a full-time innovation team, as well as implementation experts, machine learning experts, and clinical experts. All members brought years of experience solving local problems across clinical silos during which they built networks and professional credibility that were crucial to the success of the project. For several years leading up to the Sepsis Watch launch, the clinical lead reviewed sepsis cases for CMS and met with clinical stakeholders throughout the hospital to discuss sepsis treatment failures. The machine learning experts had prior experience developing machine learning models for clinical applications.

*3.4.3 Communicating Trustworthiness.* In this section, we describe four specific approaches through which the Sepsis Watch team cultivated trust and accountability in the design and development process. Sometimes this involved quantifying model performance, and sometimes this involved qualitatively communicating patient stories or providing direct experience of the tool. Different stakeholders often valued different forms of evidence and modes of communication [15].

The first approach was to focus on demonstrating progress solving a problem important to local stakeholders, rather than demonstrating progress building a novel technology. Local data was carefully curated and the solution needed to work effectively in the local context. To cultivate trust and accountability in how well the technology worked, the sepsis definition and model were peer-reviewed and disseminated in both clinical and technical venues [18, 19, 30]. The MGP-RNN framework was directly compared to both clinical scores and machine learning methods utilized to predict sepsis at other institutions. In retrospective local data, MGP-RNN consistently detected more sepsis events early than any other method.

The second approach was to communicate the utility of the innovation in ways that were meaningful to various stakeholders. Aggregate model performance measures included the number of high risk alerts per hour and the number of sepsis cases detected daily by MGP-RNN compared to other methods. Individual patient cases were also visualized to show how MGP-RNN often detected sepsis hours before the clinical diagnosis. Relevant patient cases featuring successes and failures were presented during clinical and faculty meetings to reinforce clinical experiences. A one-page "Model Facts" sheet shown in Figure 1 was designed with clinical professionals to transparently report relevant information to front-line staff. The design builds upon "Model Cards for Model Reporting" [36] and incorporates concepts from pharmaceutical drug labels.

---

**Model Facts**

**Model name:** Deep Sepsis
**Version:** 1.0

**Summary**
This model uses EHR input data collected from a patient's current encounter at the hospital to estimate the probability that the patient will meet sepsis criteria within the next 4 hours. It was developed in 2016-2019 by the Duke Institute for Health Innovation. The model was licensed to Cohere Med in July 2019.

**Mechanism**
- **Outcome** ..................................................Sepsis within the next 4 hours. See (1) for sepsis criteria
- **Output** ...................................................0% - 100% probability of sepsis occurring in the next 4 hours
- **Patient population** ..........................all adult patients >18 y.o. presenting to DUH ED and admitted
- **Time of prediction** ..............................................................every hour of a patient's encounter
- **Input data source**..................................................................Epic data extracted from Clarity
- **Input data type** ............................................demographics, analytes, vitals, medication administrations
- **Training data location and time-period** ..........................................DUH, 10/2014 – 12/2015
- **Model type**..................................................................................... Recurrent Neural Network

**Validation and performance**
- **Retrospective:** 20% random held out set from 10/2014 – 12/2015, AUROC 0.882
- **Temporal:** 6-month temporal validation set of ED visits at DUH between 03/2018 – 08/2018, AUROC 0.943
- **Prospective:** TBD
- **External:** TBD

**Uses and directions**
- **General use:** This model is intended to be used to by clinicians to identify patients for further assessment for sepsis. The model is not a diagnostic for sepsis and is not meant to guide or drive clinical care. This model is intended to complement other pieces of patient information related to sepsis as well as a physical evaluation to determine the need for sepsis treatment.
- **Examples of appropriate decisions to support:** Identification of high-risk patients for further evaluation to determine appropriateness of treatment for sepsis.
- **Before using this model:** Test the model prospectively on local data and confirm generalizability of model to the local setting and that model is effectively integrated into production system.
- **Effectiveness evaluation:** Evaluated in an emergency department to assist identification of high-risk patients for further evaluation in an emergency department. The rapid response team, nurse-driven workflow was effective at improving sepsis treatment bundle compliance.
- **Safety evaluation:** TBD

**Warnings**
- **General warnings:** This model was not trained or evaluated on patients receiving care in the ICU. Do not use this model in the ICU setting without further evaluation. This model was trained to identify the first episode of sepsis during an inpatient encounter. During long inpatient stays with multiple sepsis episodes, model accuracy needs to be further evaluated. The model is not interpretable and does not provide rationale for high risk scores. Clinical end users are expected to place model output in context with other clinical information to make final determination of diagnosis.
- **Examples of inappropriate decisions to support:** Implementation in the ICU or throughout inpatient encounters without further evaluation. Use of model as a diagnostic to guide clinical diagnosis and treatment for sepsis.
- **Discontinue use if:** Clinical staff raise concerns about how the model is being used or model performance deteriorates due to data shifts or population changes.

**Other information:**
- **Publications:**
  1) pre-print: https://www.biorxiv.org/content/10.1101/648907v1
- **Related models:**
  1) qSOFA: Seymour, C. W., Liu, V. X., Iwashyna, T. J., Brunkhorst, F. M., Rea, T. D., Scherag, A., et al. (2016). Assessment of Clinical Criteria for Sepsis: For the Third International Consensus Definitions for Sepsis and Septic Shock (Sepsis-3). *Jama*, *315*(8), 762–774. http://doi.org/10.1001/jama.2016.0288

---

**Figure 1: A "Model Facts" sheet designed to convey relevant information about the Sepsis Watch model to clinical end users. The sections of the sheet draw inspiration from pharmaceutical drug labels.**

The third approach was to establish mechanisms of public and external accountability. A clinical trial was pre-registered with specified primary and secondary outcomes. During this process, the clinical team set target goals for improvement and had candid conversations about turning off the system if improvements were not achieved. To maintain accountability, an external data safety monitoring board including top clinical researchers and the chief nursing officer was assembled to oversee the safety and efficacy of Sepsis Watch.

Finally, the fourth approach was to engage front-line clinicians directly in the design and development of the model and user interface. Clinical experts specified and reviewed representations of all data elements used by the model. Once the final model was developed, multiple iterations of chart reviews were completed to finalize the threshold used to classify high risk patients. Two RRT nurses reviewed multiple versions of the user interface and clinical experts reviewed and confirmed patient records retrospectively identified as high risk of sepsis and meeting sepsis criteria. Clinicians specified the most relevant information to accompany the risk level displayed on the Sepsis Watch user interface. This information could provide immediate insight into patient status, but needed to be contextualized and synthesized with other information.

While engaging front-line clinicians, two established practices described above were utilized to promote trust: the ability to test the innovation and the ability to watch others try the innovation first [5]. The first four users given access to Sepsis Watch during the three-month silent period preceding launch were clinical experts who helped design the system. These clinicians engaged directly with the system early on to develop intuition, provide feedback, and build experience that enhanced their ability to train others in the future.

*3.4.4 Trust and Accountability at Scale – Becoming an Enterprise Solution.* To impact patient care, Sepsis Watch needed to integrate with an EHR system purchased for $700 million that supported nearly all aspects of clinical care. It was critical that the Sepsis Watch team partner with the information technology team to enable this integration.

First, the team had to work within the network of established technology supplier relationships. Most clinical decision support is built within the EHR and building a solution outside the EHR requires significant development and integration effort. The Sepsis Watch team joined the health information technology team and EHR vendor on weekly phone calls for six months to identify an optimal path towards technical integration. Together with the EHR vendor, there was consensus that the computational requirements of the Sepsis Watch model were beyond the capabilities of the EHR. The health system invested in building interfaces to provide Sepsis Watch real-time EHR data.

At this point, the innovation team had to establish itself as a supplier of enterprise technology to the health system. Established processes, such as load testing and security reviews, were completed in the months preceding Sepsis Watch launch. Although the innovation team typically transitioned pilot projects to operational stakeholders after 12 months, the team continued to monitor and maintain the Sepsis Watch model and web application. There was agreement that "we built it, we own it". During the 6-month pilot, the Sepsis Watch team was responsive to technical issues, delivering system uptime of 99.34% with one instance of planned patching, two instances of the web application being temporarily unavailable, and one instance of data not being updated. There is also broad recognition that in healthcare concept drift and data drift can change machine learning model behavior and models have to be closely monitored to ensure reliability [12, 40, 50]. Ultimately, the innovation team set up a 55-inch monitor in the office to display the current state of the application and model risk scores, displayed in Figure 2. Clear lines of accountability were required between organizational stakeholders.

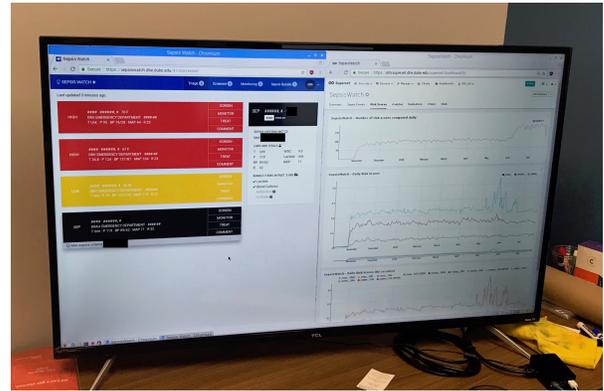

**Figure 2: A 55' monitor on a desk at the office of the innovation team. The left hand side displays the live Sepsis Watch app and the right hand side displays aggregate measures of model output.**

*3.4.5 Building Accountable Relationships with Hospital Leadership.* The two preceding sections describe how trust and accountability were cultivated to develop and then integrate Sepsis Watch at scale. This trust and accountability was channeled to develop partnerships required for the pilot. For example, when it became clear that the initial focus of the pilot was the ED, team members met with the appropriate leaders to form an implementation partnership. Two ED physicians and the ED nursing director joined the Sepsis Watch team and were closely engaged in workflow and design decisions leading up to the launch. To formalize the RRT nursing workflow, multiple team members met with the Chief Nursing Officer resulting in a subsequent meeting involving nursing leadership from across the hospital. That meeting initiated a month of close collaboration with multiple nursing stakeholders to develop training material, finalize workflow decisions, and better understand staffing requirements.

To maintain leadership engagement throughout the pilot, a governance committee was established including leadership from ED nursing, ED physicians, hospital medicine, information technology, and innovation. Monthly meetings were held throughout the pilot to review progress, address concerns surfaced by front-line staff, and make decisions about changes to workflow.

*3.4.6 Trust and Accountability on the Front Lines of Healthcare.* Significant effort was invested to build trust and accountability with RRT nurses, the primary end users of Sepsis Watch. Two nurses were involved in the design and development of the system and provided feedback on training materials developed for the pilot. During the first two weeks of the Sepsis Watch pilot, innovation team staff conducted one-on-one training sessions in the hospital to support RRT nurses during shifts. Sepsis Watch was rolled out to a small group of RRT nurses and formal and informal lines of communication were established between the nurses, clinical experts, and innovation team staff. After a month, feedback from this small group of RRT nurses was presented to the governance committee and prompted a handful of workflow changes.

To reach ED attending physicians, the Sepsis Watch team presented at multiple faculty meetings. ED physicians did not directly see the Sepsis Watch interface, so communication focused on the role of the RRT nurse and the incoming phone calls inquiring about treatment for sepsis. There was a lack of awareness during the first few weeks of the pilot and additional communication and messaging was promoted by both the ED medical and nursing directors.

Throughout the design and development of Sepsis Watch, RRT nurses and ED physicians were valued and involved as expert clinical professionals. Sepsis Watch was meant to support clinicians by ensuring that nothing slipped through the cracks. The ways in which professional expertise shaped the impact of Sepsis Watch are discussed in the next section.

## 3.5 INTEGRATION: UNEXPECTED INTERACTIONS THAT FACILITATED USE

Despite Sepsis Watch being a technical tool, its purpose was to prompt human interactions around patient care and to improve clinical decision-making. For this reason, understanding, analyzing, and improving the interactions and work practices around the tool were a central part of this project.

Based on observations of clinical practice and analyses of usage patterns by clinicians over time, we found that while early in the pilot the model may have been given too much credence, clinical professionals re-calibrated over time. Moreover, we learned that RRT nurses developed expertise and practices over time that contextualized the information displayed in Sepsis Watch, and facilitated the integration of the tool into existing clinical practice. These practices ranged from emotional labor around communication with physicians to drawing on their own clinical expertise. For instance, RRT nurses developed a practice of working outside of the app and opening a patient's EHR chart before calling the ED physician. By reading through a patient's chart, the RRT nurse was preparing to present the full clinical picture and do "due diligence," in the words of one interviewee, in anticipation of questions received from physicians. This was not a step articulated by the design team prior to implementation, but rather developed as an ad hoc practice that facilitated effective integration.

Taken together, these observations demonstrate two important findings. First, RRT nurses and ED physicians retained professional discretion to diagnose sepsis. Second, respecting the boundaries of professional practice allowed new sets of expertise to emerge and in fact enhanced the use of the machine learning-driven tool.

## 4 BEYOND INTERPRETABILITY: ALTERNATIVE STRATEGIES TO BUILT TRUST AND ACCOUNTABILITY

With reference to algorithmic fairness, accountability, and transparency, the term "black box" is often used with a negative connotation. A black box is presented as an opaque barrier that inhibits accountability [7]. However, in the healthcare context, the urge to transcend the black box is confounded by the fact that in some cases "the human body is a black box," in the words of a Sepsis Watch team member. The pre-eminent focus on machine learning model explainability or interpretability as a means to provide transparency and accountability in healthcare should be interrogated. First, front-line clinicians may not want to be oriented towards technology and away from patients. Second, the current practice of professional clinicians often includes the utilization of information that isn't comprehensively understood. Third, causal relationships are not always necessary for application in clinical decision making. Finally, as scholars have begun to point out, explainability or interpretability is poorly defined and cannot be an end in and of itself without further specification [26, 51].

What can or should constitute opening the "black box" in the context of clinical practice when the causes and mechanisms of illness, especially in the case of sepsis, are poorly understood? The response should not be to abandon the normative values driving the field toward interpretable and explainable AI, but rather should be to re-examine how those values might be achieved by other means.

In the previous section, we presented the mechanisms through which trust and accountability were built into the design, development, and implementation of Sepsis Watch. In this section, we draw out four key values and practices that should be considered when developing machine learning to support clinical decision-making: define the problem in context, build relationships with stakeholders, respect professional discretion, and create ongoing feedback loops with stakeholders.

## 4.1 PROBLEM DEFINITION: FROM END USER TO DATA SCIENTIST

As Passi and Barocas have argued, "some of the most important normative implications of data science systems have their roots in the work of problem formulation" [43]. That is, how, why, and in what context data science problems are formulated as problems to be solved is neither value-neutral nor self-obvious [39]. From this perspective, it is significant that Sepsis Watch was developed to meet a specific problem in a specific hospital as defined by frontline clinicians working in that hospital. This was achieved not only through the project selection priorities and processes, which were put in place to closely align clinical problems and new innovations, but also through the experiences of the team working on the problem. The team members had each been involved in prior attempts, some successful and some not, to implement new technologies at the local health system.

Moreover, the Sepsis Watch model was developed with clinicians from the beginning and on local patient data. Clinicians were involved in selecting the cohort parameters for the population the model was trained on, the data elements included in the model training process, and the design of the model evaluation.

*4.1.1 Existing Institutional Processes Matter.* Creating a model to improve treatment without taking into account existing care and treatment processes is likely to create not only inaccurate but also harmful health interventions [8]. This underscores the importance of working with front-line clinicians to understand relevant aspects of the care delivery process that must be included in the model as features, which Sepsis Watch did from the start.

In addition, information about the indicated and contra-indicated use of the model must be clearly articulated to end users to ensure that the strengths and weaknesses of a given model are clearly understood. The "Model Fact" sheet was designed and developed with front-line clinicians to distill relevant information about the use of the sepsis model.

## 4.2 RELATIONSHIP BUILDING: ENGAGE EARLY AND OFTEN

The multi-stakeholder development process was driven by the belief that trust in a technology is rooted in relationships—not in a technical specification or feature. Each phase of development included substantive engagement with relevant stakeholders. For instance, in the early phases of building and tuning the model, lead clinicians met regularly with the development team to review and validate the model. In the later phases involving front-line integration of the tool, the development team held regular meetings with stakeholders, from the information technology support team responsible for managing and securing EHR data to hospital management responsible for resource allocation and institutional support, to the nurses responsible for using and integrating the tool into clinical practice.

*4.2.1 Allocate Human Resources.* Defining the problem and clearly articulating how and why a technical tool addresses that problem has implications for processes around tool planning and development, as well as end-user education. The development of professional education materials around the tool's implementation was a resource-intensive aspect of the tool integration. In a formal acknowledgement of the work and resources required to integrate a tool into clinical practice, a new devoted full-time role was created to work with frontline clinicians and stakeholders. This role worked closely not only with the nursing staff responsible for using Sepsis Watch, but also acted as the liaison with the numerous hospital departments implicated in new clinical workflows, from nursing leadership to professional education modules.

## 4.3 RESPECT PROFESSIONAL DISCRETION: AUGMENT, DON'T REPLACE

The tool, from its inception, was designed to be a software diagnostic aide, not a standalone diagnostic device. This approach retained deference to human expertise to make final diagnostic decisions. The implications of relying on technology to inform decision making and diagnosis are complex. In this case, the tool was indeed designed as "an algorithm in the loop" [21]. However, there remain open research questions beyond the scope of this study around the implications for professional skill, training, and certification. While the model was not designed to improve human understanding of sepsis diagnosis, other mechanisms were in place to improve how clinicians within the ED diagnose sepsis. For example, individual patient cases of sepsis diagnosis and treatment failure were discussed at every monthly governance committee meeting. These reviews continue to be central to distilling and disseminating learnings to ED physicians. Machine learning systems need to be supplemented with other approaches to improve human understanding and decision making.

*4.3.1 Elevate New Expertise.* The RRT nurses were integral to the effective integration of Sepsis Watch, as described above. They also became experts at remotely evaluating sepsis. While nurses traditionally assess patients in person, the Sepsis Watch RRT nurses developed significant expertise contextualizing and synthesizing digital patient data. These nurses expanded their professional expertise in order to bridge the gap between digital data gathering, contextualizing, and synthesizing, and clinical decision making.

## 4.4 STAKEHOLDER FEEDBACK LOOPS: BETWEEN DESIGNERS AND USERS

The multi-stakeholder development process was designed to facilitate sustained, multi-directional communication. As much as meetings and information sessions were held to convey information to users and stakeholders, they were used equally as a site to receive information and to hold space for issue-flagging and discussion. For example, a second version of the user interface was launched two months into the pilot based on feedback received during sessions. Additionally, new kinds of feedback loops were established, such as the governance committee, to manage and oversee new issues as they arose.

In addition, Sepsis Watch developers also instantiated mechanisms to enact their commitment to ongoing monitoring of the tool in use. The tool development team embraced the mentality that, in the words of one Sepsis Watch team member, "You build it, you own it". The visual presence of Sepsis Watch in the office, with a monitor displaying the model's behavior, kept the tool from fading into the background even though they were no longer actively working on its development.

*4.3.1 Between Designers and Patient Proxies.* Although the focus of this paper has been on the clinicians impacted by Sepsis Watch, the tool was developed to improve patient outcomes and patients should not be overlooked as stakeholders in the overall system. There were several key ways in which the interests of the patients were represented. First, Sepsis Watch underwent an ethical research review process through the health system IRB, a process designed to protect ensure the safety of patient subjects. Second, Sepsis Watch was designed to be an overlay on top of existing clinical care routines, and it is within these routines that ED clinicians cultivate trust and accountability with patients. For applications of machine learning in healthcare that more directly interface with patients, additional considerations are required to cultivate trust and build mechanisms of accountability with patients.

## 5 CONCLUSION

Our presentation and discussion of the Sepsis Watch tool highlights a set of challenges and potential opportunities facing the actual, situated use of machine learning technologies in healthcare. These challenges include how to ethically build models, interpret and explain model output, recognize and account for biases, and minimize disruptions to professional expertise and work cultures. Our work underscores the necessity to conceptualize a machine learning implementation as a socio-technical system, composed of both social and technical aspects operating in a specific institutional context. This conceptualization is necessary in order to productively navigate the complex ethical dimensions of machine learning-driven healthcare.

At the outset, the goal of the Sepsis Watch tool was to improve patient care and to enhance professional expertise and decision-making. Given the lack of consensus around the clinical definition of sepsis—the ground truth the model was optimized to predict—and the potential performance benefits of a deep learning model, the team developed other mechanisms besides model interpretability to build trust and accountability in the tool's design, development, implementation, and maintenance. In our presentation of the tool, we described some of these processes and strategies, and also discussed some of the ways in which our planning did not account for the complexities of integrating and justifying the output of the model into clinical decision-making. We found that although model outputs do not need to be explainable, clinical decisions do require explanations and justifications and there is significant labor required to map between model outputs and clinical decisions. From this discussion, we drew out four key values and practices that characterized our approach to support clinical decision-making with a deep learning tool: rigorously define the problem in context, build relationships with stakeholders, respect professional discretion, and create ongoing feedback loops with stakeholders.

Our work has significant implications for future research regarding mechanisms of institutional accountability and considerations for designing machine learning systems. Our work highlights the significant investment of resources and effort to both integrate machine learning technologies and build human capacity and expertise to effectively use machine learning technologies in healthcare. Furthermore, for machine learning solutions to operate at enterprise scale, mechanisms of trust and accountability must be pursued at multiple levels within an organization. Our work underscores the limits of model interpretability as a solution to ensure transparency, accuracy, and accountability in practice. Instead, by conceptualizing Sepsis Watch as a socio-technical system in an institutional context, our work demonstrates other means and goals to achieve FATML values in design and in practice.

## ACKNOWLEDGMENTS

We wish to thank our anonymous reviewers for their insightful comments and Dina Sarro, Liz Alderton, Sahil Sandhu, and Elizabeth Watkins for their ongoing engagement and feedback. Our team would like to thank Drs. Thomas Owens, William Fulkerson, Mary Klotman, Jeffrey Ferranti, Allan Kirk, Robert Califf, and Mary Ann Fuchs for operational and leadership support throughout the development of Sepsis Watch. We would like to thank Drs. Jason Theiling, Rebecca Donahoe, Charles Gerardo, and Allan Kirk for enthusiastically supporting the initial Sepsis Watch integration in the adult emergency department. Finally, we would like to thank Kevin Anstrom, Dan Mark, and Mary Ann Fuchs for advising on the Sepsis Watch clinical study. The development, validation, and integration of Sepsis Watch was fully funded by Duke Health, with significant contribution by the Duke Institute for Health Innovation and Duke Health Technology Solutions. Data & Society's qualitative evaluation was supported by funding from Luminate, the John D. and Catherine T. MacArthur Foundation, and the Ethics and Governance of AI Fund. Funders were not involved in the preparation, review, or submission of the manuscript.